\documentclass[prd,aps,showpacs,epsf,amsfonts,
amssymb,amsmath,nofootinbib]{revtex4} %floats,

\usepackage{graphicx}

\newcommand{\be}{\begin{equation}}
\newcommand{\ee}{\end{equation}}
\newcommand{\bea}{\begin{eqnarray}}
\newcommand{\eea}{\end{eqnarray}}
\newcommand{\beaa}{\begin{eqnarray}}
\newcommand{\eeaa}{\end{eqnarray}}
\newcommand{\ba}{\begin{array}}
\newcommand{\ea}{\end{array}}
\newcommand{\bit}{\begin{itemize}}
\newcommand{\eit}{\end{itemize}}
\newcommand{\ben}{\begin{enumerate}}
\newcommand{\een}{\end{enumerate}}

\def\lab{\label}

\def\rar{\rightarrow}

\def\al{\alpha}

\def\la{\lambda}

\begin{document}

\title{Coherent states, fractals and brain
waves}

\author{Giuseppe Vitiello}

%\markboth{} {}

%\catchline{}{}{}{}{}
%%%%%%%%%%%%%%%%%%%%%%%%%%%%%%%%%%%%%%%%%%%%%%%%%%%%%%%%%%%%%%%%%%%%

%\title{\bf COHERENT STATES, FRACTALS AND BRAIN
%WAVES%\footnote{For the title, try not to use more than 3 lines.
%Typeset the title in 10 pt roman, uppercase and boldface.}
%}

%\author{GIUSEPPE VITIELLO
%\footnote{
%Typeset names n 8 pt roman, uppercase.
%Use the footnote to indicate
%the present or permanent address of the author.}
%}

\address{\it Dipartimento di Matematica e Informatica and INFN\\
Universit\`a di Salerno, I-84100 Salerno, Italy\\
vitiello@sa.infn.it - http://www.sa.infn.it/giuseppe.vitiello
%University Department, University Name, Address\\
%City, State ZIP/Zone,
%Country\footnote{State completely without abbreviations, the
%affiliation and mailing address, including country. Typeset in 8 pt
%italic.}\\
%first\_author@domain\_name
}

%\begin{history}
%\received{Day Month Year}
%\revised{Day Month Year}
%\end{history}

\begin{abstract}
I show that a functional representation of self-similarity (as the
one occurring in fractals) is provided by squeezed coherent states.
In this way,  the dissipative model of brain is shown to account for
the self-similarity in brain background activity suggested by
power-law distributions of power spectral densities of
electrocorticograms. I also briefly discuss the action-perception
cycle in the dissipative model with reference to intentionality in
terms of trajectories in the memory state space.
\end{abstract}

%\keywords{Neuronal synchronized oscillations; brain background
%activity; Self-similarity; Coherent states.
%Keyword1; keyword2;
%keyword3.
%}

\maketitle

\section{Introduction}
The first time I met Walter Freeman was in $2000$, at one of those
crowded conferences with many parallel sessions and just few (very
few!) interesting plenary talks. One of these was indeed the talk by
Lotfi Zadeh, another one was by Walter. My interest in nonlinear
dynamical systems and my curiosity in watching through the fence
what is going on in the neighboring garden brought me to that
conference. I was surprised by myself since I felt to understand a
good part of those two talks. Usually this does not happen to me. In
particular, Walter was talking of his research in mesoscopic brain
dynamics with a language very near to the one familiar to a
physicist trained in the study of the formation of ordered patterns
in condensed matter physics and in high energy physics. Of course,
he was not using the machinery of quantum field theory which is the
one used by physicists in their study of many-body systems. However,
he explained in clear words his laboratory observations and his
theoretical analysis showing that the mesoscopic neural activity of
neocortex appears to consist of the dynamical formation of spatially
extended neuronal domains in which widespread cooperation supports
brief epochs of patterned synchronized oscillations
\cite{Freeman:1975}. This vivid physical picture of the brain
mesoscopic activity was confirming to me what Hiroomi Umezawa, one
of the fathers of modern quantum field theory, was meaning by saying
that \cite{Umezawa:1995}: ''In any material in condensed matter
physics any particular information is carried by certain ordered
pattern maintained by certain long range correlation mediated by
massless quanta. It looked to me that this is the only way to
memorize some information; memory is a printed pattern of order
supported by long range correlations''. The key words here are
``widespread cooperation'' supporting ``patterned synchronized
oscillations'' and ``ordered pattern maintained by certain long
range correlation''. In 1967, when he was still in Naples, Umezawa
proposed indeed a model of brain dynamics based on the dynamical
generation of such long range correlation in the way they are
normally treated in many-body theory \cite{UR}. As a matter of fact,
there is no alternative formalism to explain the {\it dynamical}
generation of ordered patterns in solid state physics and elementary
particle physics. Crystals, magnets, and other ordered patterns
observed at reasonably high temperature, superconductors at much
lower temperature, the lowest energy state, usually called the
``vacuum'', in the standard model of particle physics, are in fact
all successfully described by many-body physics with an incredible
precision in the prediction of measured quantities. The success of
such a quantum field theory (QFT) is really impressive and led
Umezawa to say \cite{Umezawa:1995}: ''In any case soon after I moved
to Naples, I strongly felt that there should be a long range
correlation which controls the brain function. If I could know what
kind of correlation, I would be able to write down the Hamiltonian,
bringing the brain science to the level of condensed matter
physics.'' Of course, Umezawa was a physicist and long range
correlation modes were a familiar tool to work with. However,
already in the early 1940s Lashley's work showing the diffuse,
non-local nature of brain activity, induced him to talk of ``mass
action'' in the storage and retrieval of memories in the brain, and
he observed: ``...Here is the dilemma. Nerve impulses are
transmitted ...from cell to cell through definite intercellular
connections. Yet, all behavior seems to be determined by masses of
excitation...within general fields of activity, without regard to
particular nerve cells... What sort of nervous organization might be
capable of responding to a pattern of excitation without limited
specialized path of conduction? The problem is almost universal in
the activity of the nervous system'' (pp. 302--306  of Ref. 4).
%\cite{Lashley:1948}
Umezawa also knew about the work by Pribram who, soon after the
discovery of the laser light in the early 1960s
\cite{Glauber,Klauder:1968a} following the theoretical studies by
Gabor \cite{Gabor:1948,Gabor:1968}, proposed his holographic
hypothesis \cite{Pribram:1991}. The interesting point in Umezawa's
many-body model for the brain is that two main ingredients appear
there together: the notion of ``field'' introduced by Lashley in his
puzzling dilemma and the notion of ``coherence'', intrinsic to the
laser theory inspiring Pribram view. Both these notions are basic
ones in the QFT dynamics generating ordered patterns, but not in
neuroscience, and in general in biology and biochemistry, where the
atomistic view of assembling little pieces together has been
prevailing on the search of the microscopic {\it dynamical} laws
ruling their cooperative behavior so that the mesoscopic and
macroscopic functioning of the system could emerge. One must have
the courage of a Lashley and of a Pribram to dare to introduce the
field concept and the wave notion of coherence. This is why, when
listening Walter Freeman talking of dynamical widespread cooperation
supporting patterned synchronized oscillations, it was clear to me
that he is one of those few people who dare to open new paths in the
forest.

Today, with the advent of  advanced  technologies, the amplitude
modulated (AM) patterns and the phase modulated (PM) patterns in the
brain are of common observational access and it becomes imperative
to understand how out of the behavior of the single neuronal cells a
{\it transition} may occur into the coherent behavior of a
collective neuronal assembly, or in Lashley's words, ``what sort of
nervous organization might be capable of responding to a pattern of
excitation''. Laboratory observations of the brain functioning show
that the observed cortical collective activity cannot be fully
accounted for by the electric field of the extracellular dendritic
current or the extracellular magnetic field from the high-density
electric current inside the dendritic shafts, which are much too
weak, or by the chemical diffusion, which is much too slow
\cite{11,11a,FreemanNDN}. Patterns of phase-locked oscillations are
intermittently present in resting, awake subjects as well as in the
same subject actively engaged in cognitive tasks requiring
interaction with environment, so they are best described as
properties of the background activity of brains that is modulated
upon engagement with the surround. These ``packets of waves'' extend
over spatial domains covering much of the hemisphere in rabbits and
cats \cite{9,10,12,13}, and over regions of linear size of about $19
~cm$ \cite{7} in human cortex with near zero phase dispersion
\cite{14,15}. Synchronized oscillation of large-scale neuronal
assemblies in $\beta$ and $\gamma$ ranges have been detected also by
magnetoencephalographic (MEG) imaging in the resting state and in
motor task-related states of the human brain \cite{Bassett}. The AM
patterns turn out to be the result of training the subjects to
recognize stimuli under reinforcement and respond to them
appropriately. The patterns are shaped by modifications of synaptic
strengths during training. The formation of a Hebbian nerve cell
assembly thus occurs for each learned category of stimulus.

Umezawa's many-body model \cite{UR} and its extension to dissipative
dynamics, the dissipative many-body model of brain
\cite{Vitiello:1995wv} are based on the QFT notion of spontaneous
breakdown of symmetry (SBS). I want to stress that in such models
the neuron and the glia cells are treated as classically behaving
systems. The quantum degrees of freedom are related with the
symmetry of the dynamics. When such a symmetry is not the symmetry
of the least energy state (the ground state or the vacuum) of the
system, the dynamical symmetry is said to be spontaneously broken
and mathematical consistency requires the existence of massless
particles. These are called the Nambu-Goldstone (NG) quanta or modes
\cite{ITZ}. They are boson particles normally observed in solid
state physics. Examples are the phonon modes in the crystals, the
magnon modes in ferromagnets, etc. They can be described, as
customary in a quantum theory, as the quanta associated to certain
waves, such as the elastic wave in crystals, the spin wave in
ferromagnets. The role of such waves is the one of establishing long
range correlation among the system constituents. For example, in
magnets, the elementary magnetic dipoles are forced to oscillate
``in phase'' under the correlation established by the spin waves,
i.e. by the magnon quanta spanning the extended domain which thus
gets characterized by its macroscopic magnetization. We thus see
that the mechanism of spontaneous breakdown of symmetry is at the
origin of the {\it change of scale}: from the microscopic scale of
the elementary constituent dynamics to the macroscopic scale of the
system magnetization, which is therefore a measure of the ordering
of the elementary constituents and is for that reason called the
``order parameter''. An essential point is that the generation of
the ordering depends on the inner dynamics of the system, not on the
strength or on other properties of the external stimulus causing the
breakdown of the symmetry. The external field (the trigger or
stimulus) is thus responsible of the ``phase transition'' from the
normal (zero magnetization) phase to the magnetic phase. We thus
learn that the mathematical structure of the theory must be adequate
to allow physically distinct phases (technically called unitarily
inequivalent representations of the quantum algebra). QFT possesses
indeed such a mathematical structure. The  ground states
corresponding to physically distinct phases are characterized by
distinct degrees of ordering which are described by different
numbers of NG modes {\it condensed} in them. Such a condensation of
NG modes in the ground states is a {\it coherent} condensation,
which physically expresses the ``in phase'', i.e. synchronized,
dipole oscillation. In quantum mechanics (QM), on the contrary, all
the state representations are physically equivalent (unitarily
equivalent) and therefore QM is not useful to describe phase
transitions \cite{Vitiello:1998tu,Vitiello:1985}.

The quantum variables relevant to the many-body model have been
identified in subsequent developments \cite{Vitiello:1995wv,Jibu} as
the electric dipole vibrational modes of the water molecules which
constitutes the matrix in which neurons and glia cells and other
mesoscopic units are embedded. The spontaneous breakdown of the
rotational symmetry of electrical dipoles of water and other
molecules \cite{Vitiello:1995wv,Jibu,DelGiudice:1985} implies the
existence of NG modes which in such a context have been called the
dipole wave quanta (DWQ). The system ground state is obtained in
terms of coherent condensation of the DWQ. As suggested by the well
known electrical properties of cell membranes and by the
experimental observations \cite{Haider,Petersen} of slow
fluctuations in neuronal membrane polarization (the so called up and
down states) corresponding to that of spontaneous fluctuations in
the fMRI signal, the electrical dipole oscillatory matrix in which
the neuronal electrophysical and electrochemical activity is
embedded \cite{Raichle} cannot be ignored, indeed.

In the many-body model memory is obtained as an ordered pattern
produced by the coherent condensation of the DWQ under the external
stimulus acting as a trigger for the symmetry breakdown. The DWQ are
the agents by which the coordination emerges. The ``memory state''
is therefore a coherent state for the basic quantum variables, whose
mesoscopic order parameter expresses, at the synaptic level, the
amplitude and phase modulation of the carrier signal.

The original model suffers, however, of a very much limited memory
capacity. Any subsequent stimulus, producing the associated DWQ
condensation, cancels the one produced by the preceding stimulus
(overprinting). It has been shown \cite{Vitiello:1995wv} that such a
problem may find a solution when the model is modified so to include
one of the intrinsic properties of brain system, the one of being an
open system ruled, therefore, by a dissipative dynamics. The
dissipative many-body model of brain predicts then the coexistence
of physically distinct amplitude modulated and phase modulated
patterns correlated with categories of conditioned stimuli, and the
remarkably rapid onset of AM patterns into irreversible sequences
that resemble cinematographic frames. These are indeed two main
features of neurophysiological data \cite{11,11a,vortex} fitted by
the dissipative model of brain. In this paper I will not insist
further in the description of the dissipative many-body model, which
can be found in the existing literature
\cite{Vitiello:1995wv,Alfinito:2000ck,Pessa:2003,Pessa:2004,Vitiello:2001,11,11a,vortex}.
I will instead consider only two features of the model. One is
related with the scale free and self-similarity property of the
brain dynamics as suggested from power-law distributions of power
spectral densities of electrocorticograms (ECoGs). This is discussed
in Section $2$. The other one, discussed in Section $3$, is related
with the brain-environment complex coupling. Conclusive remarks are
reported in Section $4$.

\section{Self-similarity and Coherent States}
Observation of the brain background activity shows that
self-similarity characterizes the brain ground state. Indeed,
measurements of the durations, recurrence intervals and diameters of
neocortical EEG phase patterns have power-law distributions with no
detectable minima. The power spectral densities in both time and
space of ECoGs from surface arrays conform to power-law
distributions \cite{9,10,Braitenberg,Linkenkaer,Hwa}, which suggests
that the activity patterns generated by neocortical neuropil might
be scale-free \cite{Wang,Freeman2005c} with self-similarity in ECoGs
patterns over distances ranging from hypercolumns to an entire
cerebral hemisphere (this might explain the similarity of
neocortical dynamics in mammals differing in brain size by $4$
orders of magnitude, from mouse \cite{Franken} to whale
\cite{Lyamin}, which contrasts strikingly with the relatively small
range of size of avian, reptilian and dinosaur brains lacking
neocortex) \cite{11}.

In the dissipative many-body model, one of the main actors is the
(brain) ground state, which, at some initial time $t_{0}$, may be
represented as the collection (or the superposition) of a full set
of ground states, denote them by ${|0\rangle}_{\cal N}$, for all
$\cal N$. ${\cal N}$ denotes the set of integers ${\cal
N}_{a_{\kappa}}$ and ${\cal N}_{{\tilde a}_{\kappa}}$ defining the
"initial value" of the condensate, ${\cal N} \equiv \{ {\cal
N}_{a_{\kappa}} = {\cal N}_{{\tilde a}_{\kappa}}, {\rm ~for ~all}
~\kappa\}$, at  $t_{0} = 0$, and defines the {\it order parameter}
associated with the information recorded at time $t_{0} = 0$. ${\cal
N}_{a_{\kappa}}$ and ${\cal N}_{{\tilde a}_{\kappa}}$ denote the
number of DWQ $a_{\kappa}$ and ${\tilde a}_{\kappa}$ condensed in
the state ${|0\rangle}_{\cal N}$. The label $\kappa$ generically
denotes degrees of freedom such as, e.g., spatial momentum, etc. The
${\tilde a}_{\kappa}$ operators represents the system's environment;
this behaves as the time-reversed image of the dissipative system
since the energy flux {\it in-}coming into it is obtained by
time-reversal of the energy flux {\it out-}going from the
dissipative system, and vice-versa. The environment ${\tilde
a}_{\kappa}$ operators are thus time-reversed mirror modes of the
brain system. Their introduction is necessary in order to set up the
formalism for the dissipative system
\cite{Vitiello:1995wv,Celeghini:1992yv,Vitiello:2007a}. In such a
formalism the environment is thus described as the {\it
time-reversed  copy}, the {\it Double} \cite{Vitiello:2001}, of the
dissipative system. Such a {\it doubling} of the degrees of freedom
is mathematically well defined by the coproduct mapping in the
deformed Hopf algebra. Here, however, I will not insist on
mathematical details which can be found in the literature
\cite{CeleghDeMart:1995,13z}.

States ${|0\rangle}_{\cal N}$ and ${|0\rangle}_{\cal N'}$, for $\cal
N \neq N'$, are non-overlapping (unitarily inequivalent) states in
the infinite volume limit:  ${}_{\cal N}\langle 0 | 0 \rangle_{\cal
N'} \rightarrow 0~ \rm as~ {V \rightarrow \infty}$,    for all
${\cal N},~ {\cal N'} ,~~ {\cal N} \neq {\cal N'}$. The brain may
occupy any of the ground states ${|0\rangle}_{\cal N}$, or it may be
in any state that is a collection or superposition of these
brain-environment ground states. The state $|0(t)\rangle_{\cal N}$
denotes the state $| 0\rangle_{\cal N}$ at time $t$ specified by the
initial value ${\cal N}$,  at $t_{0} = 0$. ${|0 (t)\rangle}_{\cal
N}$ is recognized to be, at any time $t$, a finite temperature state
and it can be shown to be a squeezed coherent state
\cite{Vitiello:1995wv,Umezawa:1993yq,Perelomov:1986tf} (see below
for the definition of squeezing). In this Section I will show that a
functional representation of self-similarity (as the one occurring
in fractals) is provided by squeezed coherent states. In this way we
see that the dissipative model of brain accounts for the
self-similarity in brain background activity.

Below I will limit my considerations to the self-similarity property
of fractals which are generated iteratively according to a
prescribed recipe, the so-called deterministic fractals (fractals
generated by means of a random process, called ``random fractals''
\cite{Bunde}, will not be considered here. For a discussion on this
point and a relation of fractals with brain/mind see also Ref. 49.
%\cite{Stamenov}).
In some sense, the self-similarity property is the {\it most
important property} of fractals (p. 150 in Ref. 50
%\cite{Peitgen}
). The conjecture that a relation between fractals and the algebra
of coherent states exists was presented in Ref. 51
%\cite{dice2005}
. In the following I will closely follow the presentation of Ref. 52
%\cite{Licatafractals}
.

\subsection{Self-similarity and the coherent state algebra}
I consider for simplicity the {\it Koch curve} (Fig. 1). The
discussion can be extended to other iteratively constructed
fractals. Consider a one-dimensional, $d = 1$, segment $u_0$ of unit
length $L_0$, called the {\it initiator} \cite{Bunde}. As usual,
this is called the step, or stage, of order $n = 0$. The length
$L_0$ is then divided by the reducing factor $s =3$, and the
rescaled unit length $L_1 = \frac{1}{3} L_0$ is adopted to construct
the new ``deformed segment'' $u_1$ made of $\alpha = 4$ units $L_1$
(step of order $n = 1$).  $u_1$ is called the {\it generator}
\cite{Bunde}. Note that such a ``deformation'' of the $u_0$ segment
is only possible provided one ``gets out'' of the one dimensional
straight line $r$ to which the $u_0$ segment belongs: this means
that in order to construct the $u_1$ segment ``shape'' the one
dimensional constraint $d = 1$ has to be relaxed: the shape, made of
$\alpha = 4$ units $L_1$, lives in some $d \neq 1$ dimensions. Thus
we write $u_{1,q}(\alpha) \equiv q \, \alpha \, u_0$, $~~q  =
\frac{1}{3^d}$, $~~d \neq 1$, where $d$ has to be determined.

Denoting by ${\cal H}(L_0)$ lengths, surfaces or volumes, the
familiar scaling law
\be \lab{B} {\cal H}(\la L_0) = \la^{d} {\cal H}(L_0) ~, \ee
holds when lengths are (homogeneously) scaled,: $L_0 \rar \la L_0$.
A square $S$ whose side is $L_0$ scales to $\frac{1}{2^2}S$ when
$L_0 \rar \la L_0$  with $\la = \frac{1}{2}$. A cube $V$ of same
side with same rescaling of $L_0$ scales to $\frac{1}{2^3}V$. Thus
$d = 2$ and $d = 3$ for surfaces and volumes, respectively. Note
that $\frac{S(\frac{1}{2}L_0)}{S(L_0)} = p =\frac{1}{4}$ and
$\frac{V(\frac{1}{2}L_0)}{V(L_0)} = p = \frac{1}{8}$, respectively,
so that in both cases $p = \la^{d}$. For the length $L_0$, it is $p
= \frac{1}{2} = \frac{1}{2^{d}} = \la^{d}$ and of course $d = 1$.

By generalizing and extending this to the case of any other
``ipervolume'' ${\cal H}$ one considers thus the ratio
\be \lab{Ba} \frac{{\cal H}(\la L_0)}{{\cal H}(L_0)} = p ~, \ee
and assuming that Eq. (\ref{B}) is still valid ``by definition'',
one obtains
\be \lab{Bab} p ~{\cal H}(L_0) = \la^{d} {\cal H}(L_0)  ~, \ee
i.e. $p = \la^{d}$. For the Koch curve, setting $\al = \frac{1}{p} =
4$ and $q = \la^{d} = \frac{1}{3^d} $,  $p = \la^{d}$ gives
%
%\be \lab{29} L_0 = q \alpha L_0 = \frac{4}{3^d} L_0 ~, \ee
%%
%which implies
%
\be \lab{30a} q \alpha = 1~, \quad {\rm where} \quad \al = 4, \quad
q= \frac{1}{3^d}~,\ee
i.e.
\be \lab{30} d = \frac{\ln 4}{\ln 3} \approx 1.2619~. \ee
$d$ is called the {\it fractal dimension}, or the {\it
self-similarity dimension} \cite{Peitgen}.

\vspace{.6cm}

%%%%\centerline{\epsfysize=1.8truein\epsfbox{Koch0_1_2.eps}}
\vspace{.2cm}
%\begin{center}
%{\small \noindent FIG. 1. The first five stages of Koch curve.}

%%%\centerline{\epsfysize=1.8truein\epsfbox{Koch3_4.eps}} \vspace{.2cm}

%\centerline{{\small \noindent Fig. 1. The first five stages of Koch
%curve.}}

\begin{figure}
\centering \resizebox{6cm}{!}{\includegraphics{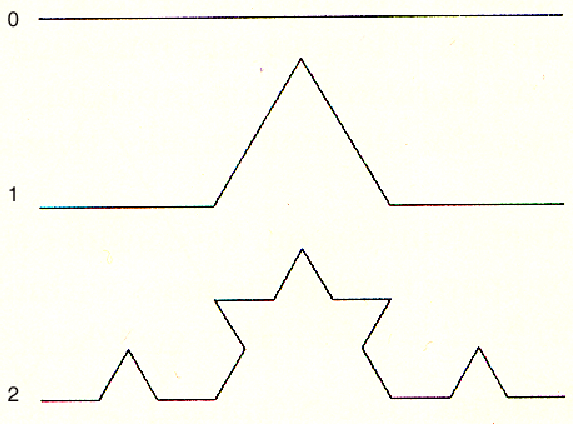}}
\centering \resizebox{6cm}{!}{\includegraphics{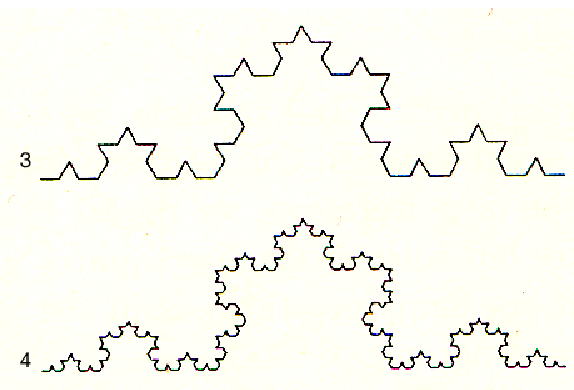}}
\caption{\small \noindent The first five stages of Koch curve.}
\label{fig1}
\end{figure}

\vspace{.6cm}

With reference to the Koch curve, I observe that the meaning of Eq.
(\ref{Bab}) is that in the ``deformed space'', to which $u_{1,q}$
belongs, the set of four segments of which $u_{1,q}$ is made
``equals'' (is equivalent to) the three segments of which $u_0$ is
made in the original ``undeformed space''. The (fractal) dimension
$d$ is the dimension of the deformed space that ensures the
existence of a solution of the relation $\frac{1}{\al} =\frac{1}{4}
= \frac{1}{3^d} = q$,  which for $d=1$ would be trivially wrong. In
this sense $d$ is a measure of the ``deformation'' of the
$u_{1,q}$-space {\it with respect to the $u_0$-space}. In other
words, we require that the measure of the deformed segment $u_{1,q}$
with respect to the undeformed segment $u_0$ be $1$:
$\frac{u_{1,q}}{u_0} = 1$, namely $\al q = \frac{4}{3^{d}} = 1$. In
the following, for brevity I will thus set $u_0  = 1$, whenever no
misunderstanding arises.

Since the deformation of $u_0$ into $u_{1,q}$ is performed by
varying the number $\al$ of rescaled unit segments from $3$ to $4$,
we expect that $\al$ and its derivative $\frac{d}{d\al}$ play a
relevant role in the fractal structure. We will see indeed that
$(\al, \frac{d}{d\al})$ play the role of conjugate variables (cf.
Eq. (\ref{(aaex2.7)})).

Steps of higher order $n$, $n = 2,3,4,..\infty$, can be obtained by
iteration of the deformation process keeping $q = \frac{1}{3^d}$ and
$\alpha = 4$ . For the $n$th order deformation we have
\be \lab{31a} u_{n,q}(\alpha) \equiv (q \, \alpha) \,
u_{n-1,q}(\alpha) ~ , \quad n = 1,2,3,... \ee
i.e., for any $n\in \mathcal{N}_+$
\be \lab{31} u_{n,q}(\alpha)  = (q \, \alpha)^{n} \, u_0 ~ . \ee

By proceeding by iteration, %in Eq. (\ref{29}),
or, equivalently, by requiring that $\frac{u_{n,q}(\alpha)}{u_0}$ be
$1$ for any $n$, gives $(q \, \alpha)^{n} = 1$ and Eq. (\ref{30}) is
again obtained.  Notice that the fractal is mathematically defined
in the limit $n \rar \infty$ of the deformation process. The
definition of fractal dimension is indeed given starting from $(q
\al)^n =1$ in the $n \rar \infty$ limit. Since $L_n \rar 0$ for $n
\rar \infty$, the Koch fractal is a curve which is everywhere
non-differentiable \cite{Peitgen}.

Eqs. (\ref{31a}) and (\ref{31}) express in analytic form, {\it in
the $n \rar \infty$ limit}, the {\it self-similarity} property of a
large class of fractals (the Sierpinski gasket and carpet, the
Cantor set, etc.): ``cutting a piece of a fractal and magnifying it
isotropically to the size of the original, both the original and the
magnification look the same'' \cite{Bunde}. In this sense one also
says that fractals are ``scale free'', namely viewing a picture of
part of a fractal one cannot deduce its actual size if the unit of
measure is not given in the same picture \cite{BakinBunde}. I stress
that only in the $n \rar \infty$ limit self-similarity is defined
(self-similarity does not hold when considering only a finite number
$n$ of iterations).

I recall that invariance,  in the limit of $n \rar \infty$
iterations, only under anisotropic magnification is called
self-affinity. The discussion below can be extended to self-affine
fractals. I will not discuss here the measure of lengths in
fractals, the Hausdorff measure, the fractal ``mass'' and other
fractal properties. The reader is referred to the existing
literature.

My main observation is now that Eq. (\ref{31}) expresses the deep
formal connection with the theory of coherent states and the related
algebraic structure. In order to make such a connection explicit, I
observe that, by considering in full generality the complex
$\al$-plane, the functions
\be \lab{(a2.8)} u_n(\al) = {\al^n\over \sqrt{n!}} ~,~\quad u_0
(\al) = 1~, ~~ \quad \quad~~ n\in \mathcal{N}_+ ~, \quad \al \in
%\mathds
{\cal C} ~,
%\mlab{(2.8)}
\ee
form in the space ${\cal F}$ of the entire analytic functions a
basis which is orthonormal under the gaussian measure $d\mu(\al)=$
$\displaystyle{{1\over{\pi}} e^{- |\al|^2} d\al d{\bar \al}}$. In
Eq. (\ref{(a2.8)}) the factor $\frac{1}{\sqrt{n!}}$ ensures the
normalization condition with respect to the gaussian measure. In the
following it is always $n\in \mathcal{N}_+$.

The functions $u_{n,q}(\alpha)|_{q \rar 1}$ in Eq. (\ref{31}) are
thus immediately recognized to be nothing but the restriction to
real $\al$ of the functions in Eq. (\ref{(a2.8)}), apart the
normalization factor $\frac{1}{\sqrt{n!}}$. The study of the fractal
properties may be thus carried on in the space ${\cal F}$ of the
entire analytic functions, by restricting, at the end, the
conclusions to real $\al$, $\al \rar {\it Re}(\al)$. Furthermore,
since actually in Eq. (\ref{31}) it is $q \neq 1$ ($q < 1$), one
also needs to consider the ``$q$-deformed'' algebraic structure of
which the space ${\cal F}$ provides a representation.

The space ${\cal F}$  is a vector space which provides the so called
Fock-Bargmann representation (FBR) \cite{Perelomov:1986tf,Fock} of
the Weyl--Heisenberg algebra generated by the set of operators $\{
a, a^{\dagger}, {\bf 1} \}$:
\be \lab{(aex2.7)} [a, \,a^\dagger] = {\bf 1}~, \qquad [N,
\,a^\dagger] = a^\dagger~, \qquad [N, \,a] = - a ~, \ee
where $N \equiv a^{\dagger} a$, with the identification:
\be \lab{(aaex2.7)} N \to \al {d\over d\al} ~ ,~\quad a^\dagger \to
\al ~ ,~\quad a \to
{d\over d \al} ~ . %\mlab{(ex2.7)}
\ee

The $u_n(\al)$ (Eq. (\ref{(a2.8)})) are eigenkets of $N$ with
integer (positive and zero) eigenvalues. The FBR is the Hilbert
space ${\cal K}$ generated by the $u_n(\al)$, i.e. the whole space
${\cal F}$ of entire analytic functions. Any vector
$\displaystyle{|\psi \rangle}$ in ${\cal K}$ is associated, in a
one-to-one correspondence, with a function $\psi (\al) \in {\cal F}$
and is thus described by the set $\{c_n ; ~c_n\in \mathcal{C},
~\sum_{n=0}^\infty |c_n|^2 = 1 \}$ defined by its expansion in the
complete orthonormal set of eigenkets $\{ |n\rangle \}$ of $N$:
\bea \lab{(aex2.9)} |\psi \rangle  = \sum_{n=0}^\infty c_n |n\rangle
&\rightarrow&  \psi (\al) = \sum_{n=0}^\infty c_n u_n(\al), \\
\lab{(aex2.9c)} \langle \psi|\psi \rangle =  \sum_{n=0}^\infty
|c_n|^2 &=& \int |\psi (\al)|^{2} d \mu (\al) = ||\psi||^{2}=1, \\
\lab{(aex2.9b)} |n\rangle  &=& \frac{1}{\sqrt{n!}} (a^\dag)^{n}| 0
\rangle ~, \eea
where $|0\rangle$ denotes the vacuum vector, $a |0\rangle = 0$,
$\langle 0|0 \rangle = 1$. The series expressing $\psi (\al)$ in Eq.
(\ref{(aex2.9)}) converges uniformly in any compact domain of the
$\al$-plane due to the condition $\sum_{n=0}^\infty |c_n|^2 = 1$
(cf. Eq. (\ref{(aex2.9c)})), confirming that $\psi (\al)$ is an
entire analytic function.

The Fock--Bargmann representation provides a simple frame to
describe the usual coherent states (CS)
\cite{Perelomov:1986tf,Klauder:1968a} $|\alpha \rangle$:
\be \lab{(3.1)} |\alpha\rangle = {\cal D}(\alpha ) |0\rangle ~
,\quad a |\alpha\rangle = \alpha |\alpha\rangle~ ,  \quad \alpha \in
%\mathds
{\cal C} ~, \ee
\be \lab{(3.2)} |\alpha\rangle = \exp\biggl(- {|\alpha|^2\over 2}
\biggr) \sum_{n=0}^\infty {{\alpha ^n}\over {\sqrt{n!}}} |n\rangle =
\exp\biggl(-{{|\alpha|^2}\over 2}\biggr) \sum_{n=0}^\infty
u_n(\alpha) |n\rangle .  \ee
The unitary displacement operator ${\cal D}(\alpha)$ in
(\ref{(3.1)}) is given by:
\be \lab{(3.3)} {\cal D}(\alpha) = \exp\bigl(\alpha a^\dagger -{\bar
\alpha} a \bigr) = \exp\biggl(-\frac{|\alpha|^2}{2}\biggr)
\exp\bigl(\alpha a^\dagger\bigr) \exp\bigl(-{\bar \alpha} ~a\bigr)
~. \ee
${\cal D}(\alpha)$ is a bounded operator defined on the whole ${\cal
K}$. It provides a representation of the Weyl--Heisenberg group
usually denoted by $W_{1}$ \cite{Perelomov:1986tf}. We have
\be \lab{(3.4a)} ~~{\cal D}^{-1}(\alpha)~a~{\cal D}(\alpha) = ~a ~+
~ \al ~. \ee

The explicit relation between the CS and the entire analytic
function basis ~$\{ u_n(\al) \}$ is:
\be \lab{(3.2a)} u_n (\alpha) = {\rm e}^{{1\over 2}|\alpha|^2}
\langle n|\alpha \rangle~. \ee
%

%The following relations hold
%%
%\bea \lab{(3.4a)} ~~{\cal D}^{-1}(\alpha)&a&{\cal D}(\alpha) = ~a ~+
%~ \al ~,\\
%\lab{(3.4)} {\cal D}(\alpha) {\cal D}(\beta) &=& \exp\bigl( i
%{\it Im}(\alpha {\bar \beta})\bigr) {\cal D}(\alpha + \beta)~ , \\
%\lab{(3.5)} {\cal D}(\alpha) {\cal D}(\beta) &=&\exp\bigl( 2 i {\it
%Im}(\alpha {\bar \beta})\bigr) {\cal D}(\beta) {\cal D}(\alpha) ~.
%\eea
%%

 The set $\{ |\alpha \rangle \}$
is an overcomplete set of states, from which, however, a complete set can be
extracted. Is well known that in order to extract a complete set of CS from the
overcomplete set it is necessary to introduce in the $\alpha$-complex plane a regular
lattice $L$, called the von Neumann lattice \cite{Perelomov:1986tf}.  For a general
discussion and original references see Ref. 47.
%\cite{Perelomov:1986tf}
See also Refs. 44
%\cite{CeleghDeMart:1995}
where the von Neumann lattice is
discussed also in connection with the deformation of the
Weyl-Heisenberg algebra introduced below.

I now introduce the finite difference operator ${\cal D}_q$ defined
by \cite{13z}:
\be \lab{(2.12)} {\cal D}_q f(\al) = {{f(q \al) - f(\al)}\over
{(q-1) \al}} ~, \ee
with ~$f(\al) \in {\cal F}\; ,\; q = e^\zeta \; ,\; \zeta \in
{%\mathds
{\cal C}}$ . ${\cal D}_q$ reduces to the standard derivative for $q
\to 1$ ($\zeta \to 0$). In the space ${\cal F}$, ${\cal D}_q$
satisfies, together with $\al$ and $\al {d\over {d\al}}$, the
commutation relations:
\be \lab{(2.17)} \bigl[ {\cal D}_q , \al \bigr] = q^{\al {d\over
{d\al}}} \quad ,\quad \left [ \al {d\over d\al} , {\cal D}_q \right
] = - {\cal D}_q \quad ,\quad \left [\al {d\over d\al} , \al \right
] = \al ~, \ee
which, as for Eq. (\ref{(aaex2.7)}), lead us to the  identification
\be \lab{(2.18)} N \to \al {d\over d\al} \quad ,\quad {\hat a}_q \to
\al \quad ,\quad a_q \to {\cal D}_q ~, \ee
with ~${\hat a}_q = {\hat a}_{q=1} = a^\dagger$~ and ~$\lim_{q\to 1}
a_q = a$ on ${\cal F}$. The algebra (\ref{(2.17)}) is the
$q$-deformation of the algebra (\ref{(aex2.7)}). For shortness I
omit to discuss further the properties of ${\cal D}_q $ and the
$q$-deformed algebra (\ref{(2.17)}). More details can be found in
Refs. 44.
%\cite{CeleghDeMart:1995}
Here I only recall that the operator $q^N$ acts on the whole ${\cal
F}$ as
\be \lab{(2.20)}  q^{N} f(\al) = f(q \al) ~, \quad f(\al) \in {\cal
F}~. \ee
This result was originally obtained in Refs. 44,
%\cite{CeleghDeMart:1995}
where it was realized that the $q$-deformation arises whenever one
deals with some finite scale. A finite scale occurs indeed also in
the present case of fractals and therefore also in this case we in
fact have a deformation of the algebra.

Eq. (\ref{(2.20)}) applied to the coherent state functional
(\ref{(3.2)}) gives
\be \lab{(a2.21)}  q^{N} |\al \rangle = |q \al \rangle =
\exp\biggl(- {{|q\alpha|^2}\over 2}\biggr) \sum_{n=0}^\infty
\frac{(q \alpha)^{n}}{\sqrt{n!}}~ |n\rangle~, \ee
and, since $q \al \in %\mathds
{\cal C}$, from Eq. (\ref{(3.1)}),
\be \lab{(ab2.21)}  a~ |q \al \rangle = q \al ~|q \al \rangle ~,
\quad q \al \in %\mathds
{\cal C}~. \ee
By recalling that we have set $u_0 \equiv 1$, the $n$th fractal
iteration, Eq. (\ref{31}), is obtained by projecting out the $n$th
component of $|q \al \rangle$ and restricting to real $q \al$, $q
\al \rar {\it Re} (q \al)$:
\be \lab{(a2.22)} u_{n,q} (\al) = (q \alpha)^{n} = {\sqrt{n!}}~
\exp\biggl({{|q\alpha|^2}\over 2}\biggr) \langle n|q \al \rangle, ~~
 ~ q \al \rar {\it Re} (q
\al).\ee
for any  $n\in \mathcal{N}_+$. Taking into account that $\langle n|
= \langle 0| ~\frac{(a)^{n}}{\sqrt{n!}}$,  Eq. (\ref{(a2.22)}) gives
\be \lab{(a2.23)} u_{n,q} (\al) = (q \alpha)^{n} =
\exp\biggl({{|q\alpha|^2}\over 2}\biggr) \langle 0|(a)^{n}|q \al
\rangle, ~~~ n\in \mathcal{N}_+, ~~~ q \al \rar {\it Re} (q \al),
\ee
which shows that the operator $(a)^{n}$ acts as a ``magnifying''
lens \cite{Bunde}: the $n$th iteration of the fractal can be
``seen'' by applying $(a)^{n}$ to $|q \al \rangle$ and restricting
to real $q \al$:
\be \lab{(ab2.24)}  \langle q \al | (a)^{n} |q \al \rangle =  (q
\alpha)^{n} = u_{n,q} (\al), \quad q \al \rar {\it Re} (q \al). \ee

Eq. (\ref{(ab2.21)})  expresses the invariance of the coherent state
representing the fractal under the action of the operator
$\frac{1}{q \al}a$~. This reminds us of the fixed point equation
$W(A) = A$, where $W$ is the Hutchinson operator \cite{Bunde},
%{Hutchinson in Bundep171},
characterizing the iteration process for the fractal $A$ in the $n
\rar \infty$ limit. Such an invariance property allows to consider
the coherent functional $\psi (q\al)$ as an ``attractor'' in
$%\mathds
{\cal C}$.

In conclusion, the operator $q^{N}$ applied to $|\al \rangle$
``produces'' the fractal in the functional form of the coherent
state $|q \al \rangle$ (cf Eq. (\ref{(a2.21)})). The $n$th fractal
stage of iteration, $n = 0,1,2,..,\infty$ is represented, in a
one-to-one correspondence, by the $n$th term in the coherent state
series in Eq. (\ref{(a2.21)}). I call $q^{N}$ {\it the fractal
operator}.

Eqs. (\ref{(a2.22)}), (\ref{(a2.23)}) and (\ref{(ab2.24)}) formally
establish the searched connection between fractal self-similarity
and the ($q$-deformed) algebra of the coherent states.

\subsection{Self-similarity and squeezed coherent states}
I now look at the fractal operator $q^N$ from a different
perspective and consider the identity
\be \lab{(4.1)} 2 \al {d\over {d\al}} \psi (\al) = \left\{{1\over 2}
\left[\left(\al + {d\over {d\al}}\right)^2 - \left(\al - {d\over
{d\al}}\right)^2\right] - 1\right\} \psi (\al) ~ , \ee
which holds in the Hilbert space identified with the space ~${\cal
F}$ of entire analytic functions $\psi (\al)$. It is convenient to
set $\al \equiv x + iy$, $x$ and $y$ denoting the real and the
imaginary part of $\al$, respectively. I then introduce the
operators
\be \lab{(4.2a)} {c} = {1\over {\sqrt{2}}} \bigl( \al +  {d\over
{d\al}} \bigr) \quad ,\quad { c}^\dagger = {1\over {\sqrt{2}}}
\bigl( \al - {d\over {d\al}}\bigr)\quad , \quad [c, c^\dagger ] =
%\mathds
{\bf 1} ~. \ee
%
%Their relation with the FBR operators $a$ and $a^\dagger$ is
%%
%\be \lab{(4.2b)} \al = {1\over {\sqrt{2}}} ( c + {c}^\dagger )~ \to
%a^\dagger ~ , \quad {d\over {d\al}} = {1\over {\sqrt{2}}} ( c -
%{c}^\dagger ) ~\to a ~.  \ee
%%

In ~${\cal F}$, $~{c}^\dagger$ is the conjugate of $c$
\cite{Perelomov:1986tf,CeleghDeMart:1995}. In the limit $\al \rar
{\it Re} (\al)$, i.e.  $y \to 0$,~ $c$ and $c^\dagger$ turn into the
conventional annihilation and creator operators associated with $x$
and $p_x$ in the canonical configuration representation,
respectively. I now remark that the fractal operator $q^N$ can be
realized in  ~${\cal F}$ as:
\be \lab{(4.3)} q^N \psi (\al) = {1\over{\sqrt q}}~\exp\biggl({
\zeta \over 2}\bigl(c^2 - {c^\dagger}^2\bigr)\biggr)\psi (\al)
\equiv {1\over{\sqrt q}} {\hat {\cal S}(\zeta)} \psi (\al) \equiv
{1\over{\sqrt q}} \psi_{s}(\al) ~,  \ee
where $q = e^{\zeta}$ (for simplicity, assumed to be real) and as
usual $N = \al {d\over {d \al}}$.  ${\hat {\cal S}(\zeta)}$ is
defined to be the squeezing operator  well known in quantum optics
\cite{CeleghDeMart:1995,Yuen:1976vy}. $\zeta = \ln q $ is called the
squeezing parameter. In (\ref{(4.3)}) $\psi _s(\al)$ denotes the
squeezed states in FBR. From Eq. (\ref{(4.3)}) we see that $q^N$
acts in ~${\cal F}$, as well as in the configuration representation
in the limit $y \to 0$, as the squeezing operator ${\hat {\cal
S}(\zeta)}$, up to the numerical factor ${1\over{\sqrt q}}$.

Since $ q^N \psi (\al) =  \psi (q \al)$ (cf. Eq. (\ref{(2.20)})),
from Eq. (\ref{(4.3)}) we see that the $q$-deformation process,
which we have seen is associated to the fractal generation process,
is equivalent to the squeezing transformation.

The right hand side of (\ref{(4.3)}) is a $SU(1,1)$ group element.
%In fact, by defining $K_{-} = {1\over 2}c ^2$, $K_{+} = {1\over
%2}c^{\dagger 2}$, $K_{\al} = {1\over 2}(c^\dagger c + {1\over 2})$,
%one easily checks they close the algebra $su(1,1)$.
We indeed obtain the $SU(1,1)$ Bogoliubov (squeezing)
transformations for the $c$'s operators:
\bea \lab{(4.4b)} {\hat {\cal S}}^{-1}(\zeta) ~c~ {\hat {\cal
S}}(\zeta) = c ~\cosh \zeta - {c}^\dagger ~\sinh \zeta ~,\\
{\hat {\cal S}}^{-1}(\zeta) ~{c}^\dagger ~{\hat {\cal S}}(\zeta) =
{c}^\dagger ~\cosh \zeta -  c ~\sinh \zeta ~, \eea
and in the $y \to 0$ limit
\bea \lab{(4.4d)} {\hat {\cal S}}^{-1}(\zeta)~ \al~ {\hat {\cal
S}}(\zeta) &=& {1\over{q}}\al \rar {1\over{q}}x ~ , \\
\lab{(4.4e)}{\hat {\cal S}}^{-1}(\zeta)~p_{\al} ~{\hat {\cal
S}}(\zeta) &=& q p_{\al} \to qp_x ~ ,  \eea
where ${p_\al \equiv -i{d\over{d\al}}}$, and
\bea \lab{(4.5)} \int d\mu(\al) {\bar \psi}(\al){\hat {\cal S}}^{-
1}(\zeta)~ \al ~{\hat {\cal S}} (\zeta) \psi (\al)   &\rar&
{1\over{q}}<x>\\
 \int d\mu(\al) {\bar \psi}(\al){\hat {\cal S}}^{- 1}(\zeta)~
 p_{\al}~
{\hat {\cal S}} (\zeta) \psi (\al)   &\rar&  q<p_{x}>  \eea
so that the root mean square deviations $\Delta x$ and $\Delta
p_{x}$ satisfy
\be \lab{(4.6)} \Delta x \Delta p_{x} = {1\over 2} \quad ,\quad
\Delta{x} = {1\over q} {\sqrt{1\over 2}}\quad ,\quad \Delta {p_{x}}
= q{\sqrt{1\over 2}}~ . \ee
This confirms that the $q$-deformation plays the role of squeezing
transformation. Note that the action variable $\int p_{x}~dx$ is
invariant under the squeezing transformation.

Eq. (\ref{(4.4d)}) shows that $\al \rar {1\over{q}}\al$ under
squeezing transformation, which, in view of the fact that $q^{-1} =
\al$ (cf. Eq. (\ref{30a})), means that $\al \rar \al^{2}$, i.e.
under squeezing we proceed further in the fractal iteration process.
Thus, the fractal iteration process can be described in terms of the
coherent state squeezing transformation.

Besides the scale parameter one might also consider, phase
parameters and translation parameters characterizing (generalized)
coherent states (such as $SU(2)$, $SU(1,1)$, etc. coherent states).
For example, by changing the parameters in a {\it deterministic
iterated function process}, also referred to as {\it multiple
reproduction copy machine} process \cite{Peitgen}, (such as phases,
translations, etc.) the Koch curve may be transformed into another
fractal (e.g. into Barnsley's fern \cite{Peitgen}). In the scheme
here presented, these fractals are described by corresponding
unitarily inequivalent representations in the limit of infinitely
many degrees of freedom (infinite volume limit)
\cite{Licatafractals}. See Ref. 52
%\cite{Licatafractals}
for further
details on the functional representation of self-similarity in terms
of entire analytic functions.

I conclude that, since the vacuum states in the dissipative
many-body model are squeezed coherent states, they provide the
functional representation of self-similarity observed in
neuro-phenomenological data.

\setcounter{equation}{0}

\section{Trajectories in the Attractor Landscape}
Dissipation is a key ingredient which allows to exploit the
infinitely many unitarily inequivalent representations of QFT. Each
spatial AM pattern is described to be consequent to spontaneous
breakdown of symmetry triggered by external stimulus and is
associated with one of the unitarily inequivalent ground states.
Their sequencing is associated to the non-unitary time evolution
implied by dissipation. Changes in the brain--environment
interaction produce changes in the brain ground state. The brain
evolution through the vacuum states thus reflects the evolution of
the coupling of the brain with the surrounding world. In the memory
space, i.e. the {\it brain state space},  each representation
$\{{|0\rangle}_{\cal N}\}$ denotes a physical phase of the system
and may be conceived as a ``point'' identified by a specific $\cal
N$-set in a ``landscape of attractors''. Vacuum states are indeed
least energy states towards which time evolution proceeds and thus
they act as  dynamical ``attractors''.  Under the influence of one
or more stimuli brain may undergo an extremely rich sequence of
phase transitions, namely a sequence of dissipative structures
formed by AM patterns, through trajectories in such landscape of
attractors.

These trajectories  turn out to be {\it classical} trajectories
\cite{Vitiello:2003me} which may also be {\it chaotic}
\cite{Pessa:2003,Pessa:2004} and {\it itinerant} through a chain of
'attractor ruins' \cite{Tsuda,Skarda,Kelso,Bressler,Fingelkurts}.
{\it The possibility of deriving from the microscopic dynamics the
classicality of such trajectories is one of the merits of the
dissipative many-body model}.

The entropy, for both $a$ and $\tilde a$ system, is found to grow
monotonically from $0$ to infinity as the time goes   to $t =
\infty$ \cite{Vitiello:1995wv}. For the complete system $a-{\tilde
a}$, the difference ~$(S_{a} - S_{\tilde a})$~ is constant in time.
%: $ [\,S_{a} - S_{\tilde a} , {\cal H}^{\prime} ] = 0$.
The change in the energy $ {E_{a} \equiv \sum_{k} E_{k} {\cal
N}_{a_{k}}}$  and in the entropy is given by
\be d E_{a} = \sum_{k} E_{k} \dot{\cal N}_{a_{k}} d t =
 {1\over{\beta}} d {\cal S}_{a}  ~ ,
  \lab{(23)}
\ee
so that the free energy ${\cal F}_{a}$ of the brain system is
minimized on the trajectories:
\be d {\cal F}_{a} \equiv d E_{a} - {1\over{\beta}} d {\cal S}_{a} =
0 ~, \lab{(24)} \ee
provided changes in inverse temperature are slow, i.e. $ {{{\partial
\beta}\over{\partial t}} = - {1\over{k_{\tilde a} T^{2}}} {{\partial T}\over{\partial
t}} \approx ~0}$, which is what actually happens in mammalian brains which keep their
temperature nearly constant. As usual heat is defined as $ {dQ={1\over{\beta}} dS}$.
The change in time of condensate  ($\dot{\cal N}_{a_{k}}$, Eq. (\ref{(23)})) turns
out to be heat dissipation $dQ$. Thus, entropy changes and heat dissipation involved
in the disappearance/emergence of the coherence (ordering) associated to the AM
patterns turns into energy changes. Heat dissipation is indeed a significant variable
in laboratory observations. Brains require constant perfusion with arterial blood and
venous removal to dispose of waste heat \cite{vortex}. I observe that entropy
variation of the system also implies variation of the entropy  of the environment.
The reciprocal system--environment interaction is thus a reciprocal back--reaction
process. Then, the process of minimizing the free energy characterizing the system
evolution can be thought as a "survival" strategy of the system producing continual
adaptation of system to the environment and, at once, continual environmental
modifications.

The model predicts condensation domains of different finite sizes
with different degrees of stability \cite{Alfinito:2000ck}.
%The
%condensation function $f(x)$ which acts as a ``form factor" specific
%for the considered domain
%\cite{Umezawa:1993yq,Alfinito:2002a,Alfinito:2002b} plays a crucial
%role. $f(x)$ has to carry some topological singularity in order for
%the condensation process to be physically detectable. Similarly,
%%A regular
%%function $f(x)$ would produce a condensation which could be easily
%%``washed" out (``gauged" away by a convenient gauge transformation).
%the phase transition between inequivalent spaces can only be induced by  a singular
%$f(x)$. This  is why topologically non trivial extended objects, such as vortices,
%appear in phase transitions
%\cite{Umezawa:1993yq,Alfinito:2002a,Alfinito:2002b,vortex}.
Phase transitions driven by boson condensation are associated with some singularity
in the field phase at the phase transition point
\cite{11a,vortex,Alfinito:2002a,Alfinito:2002b}. This specific feature of the model
accounts \cite{vortex} for a crucial mechanism observed in laboratory experiments:
the event that initiates a perceptual phase transition is an abrupt decrease in the
analytic power of the background activity to near zero (null spike), associated with
the concomitant increase of spatial variance of analytic phase. The null spikes recur
aperiodically at rates in the theta  ($3-7 ~Hz$) and alpha ($8-12 ~Hz$) ranges. These
null spikes are of crucial importance for the brain-environment interaction. They
allow the readiness of the brain to react to an  input incoming at or just before the
null spike.

The incoming input is ``expected'' on the basis of the attractor
landscape as modified by the brain ``experiences'' with previously
received inputs. In this way, on the basis of these recurrent {\it
rearrangements} of the attractor landscapes, the brain promotes and
controls the {\it action} which is thus {\it intentionally} guided
in order to have the ``maximal grip'' \cite{MerleauP,Dreyfus} on the
external world. The selective sensitivity to perceptual inputs which
better fit the existing attractor landscapes (which specifies in
which sense incoming inputs are ``expected'') is controlled by the
process of activation of mesoscopic neural patterns related with
such landscapes and is termed preafference \cite{Kay,Kozma}.

Moreover, in the (cyclic) process of action-perception, the
continual reshaping and rearrangement of the attractor landscapes,
due to the introduction in the ``memory state space'' of the new
vacuum condensate triggered by a forthcoming stimulus, constitutes
the ``contextualization'' process by which, by differentiation with
preexisting landscape arrangement, a {\it meaning}
\cite{Jornten,Rogers} is attached to such incoming stimulus. The
experience of the brain in its relation with the world in which it
is {\it situated} thus becomes ``stored experience'', or {\it
knowledge} \cite{11,11a}, which generates perspectives, hypothesis,
expectations, i.e. the {\it view of the world} reached by that brain
in its experience history \cite{Vitiello:2001,Vitiello:2008atq}. Any
subsequent action becomes then a {\it test} for such a ``view of the
world'', which is trustable and generates ``confidence'' exactly
because it cannot escape from being continuously tested in the
un-resting action-perception cycle (the brain is intrinsically and
thus permanently open on the world). The process of knowledge as
described above thus implies in an essential way the rearrangement
of the attractor landscape (contextualization),  not just
``additions'' of new ``points'' in the landscape.

It is interesting that the picture describing the brain's {\it
to-be-in-the-world} \cite{MerleauP,Dreyfus}, emerging from
neuroscience observations and their dissipative model formalization
as deeply rooted in the brain un-avoidable experiential dimension,
closely depicts Galileo's paradigm for the New Science.

\section{Conclusive remarks. To-be-in-the-world: I and My Double}
As a conclusive remark, let me go back to what said on the
environment description at the beginning of Section $2$. There we
have seen that the environment is described in the dissipative
many-body model as the system time-reversed copy, its {\it Double}.
This represents the environment {\it as seen} from the system. The
inter-relation between the system (brain) and the environment is
thus the relation of the system with its Double. What we have said
in the previous Section may be rephrased in terms of a {\it dialog}
with the Double \cite{Vitiello:1995wv,Vitiello:2001}, which is a
dynamical one since there is a continuous reciprocal updating,
expressing the system's dynamical view of the world. Such a dialog
also implies the {\it emotion} of {\it novelty} and sometimes the
{\it surprise} or {\it astonishment} of a freshly acquired {\it
different}, perhaps even {\it completely different}, view of the
world \cite{Vitiello:2001,Vitiello:2008atq,Vitiello:2004}.
Remarkably, it has been proposed \cite{Desideri:2007} that the
aesthetical experience consists in realizing the perfect
accomplishment of our trade with the external world leading us to
such an emotion of novelty \cite{Vitiello:2008atq}, which at same
time was searched and even expected on the basis of our pre-existing
landscape scenarios (``education'' through the ``acquired
experiences''). This may shed some light on the relation between the
aesthetical experience and Spinoza's ``intuitive science''
\cite{Diodato:1997}.

To-be-in-the-world thus becomes this life ``entre-deux'', however
diffused in the world, the {\it between} expressing the reciprocal
being {\it exposed} to each other sight, in a continual trade
\cite{Vitiello:2004}. Consciousness mechanisms may have their roots
in this dialog with the Double
\cite{Vitiello:1995wv,Vitiello:2001,Vitiello:2004}. Let me then
close the paper by reporting the following ``thoughts''
\cite{Borges}:

\vspace{1,5mm}

``{\it The other one, the one called Borges, is the one things
happen to....It would be an exaggeration to say that ours is a
hostile relationship; I live, let myself go on living, so that
Borges may contrive his literature, and this literature justifies
me....Besides, I am destined to perish, definitively, and only some
instant of myself can survive him....Spinoza knew that all things
long to persist in their being; the stone eternally wants to be a
stone and a tiger a tiger. I shall remain in Borges, not in myself
(if it is true that I am someone)....Years ago I tried to free
myself from him and went from the mythologies of the suburbs to the
games with time and infinity, but those games belong to Borges now
and I shall have to imagine other things. Thus my life is a flight
and I lose everything and everything belongs to oblivion, or to him.

I do not know which of us has written this page.}''

\section*{Acknowledgments}

Financial support from INFN and University of Salerno  is
acknowledged. The author is grateful to M. Blasone and P.Jizba for
useful discussions on fractals.

\end{document}